\documentclass[10pt,twocolumn,english,aps,pra,superscriptaddress,address,showpacs,showacknowledgments]{revtex4-1}
\usepackage[T1]{fontenc}
\usepackage[latin9]{inputenc}
\usepackage[a4paper]{geometry}
\usepackage[active]{srcltx}
\usepackage{xcolor}
\usepackage{amsmath}
\usepackage{amssymb}
\usepackage{stackrel}
\usepackage{graphicx}
\PassOptionsToPackage{normalem}{ulem}
\usepackage{ulem}



\usepackage{babel}
\begin{document}

\title{Dissipative entanglement preparation via Rydberg antiblockade and
Lyapunov control}

\author{Zong-Xing Ding}

\affiliation{Fujian Key Laboratory of Quantum Information and Quantum Optics and
Department of Physics, Fuzhou University, Fuzhou, Fujian 350116, China}

\author{Chang-Sheng Hu}

\affiliation{Fujian Key Laboratory of Quantum Information and Quantum Optics and
Department of Physics, Fuzhou University, Fuzhou, Fujian 350116, China}

\author{Li-Tuo Shen}

\affiliation{Fujian Key Laboratory of Quantum Information and Quantum Optics and
Department of Physics, Fuzhou University, Fuzhou, Fujian 350116, China}

\author{Zhen-Biao Yang}

\affiliation{Fujian Key Laboratory of Quantum Information and Quantum Optics and
Department of Physics, Fuzhou University, Fuzhou, Fujian 350116, China}

\author{Huaizhi Wu}

\affiliation{Fujian Key Laboratory of Quantum Information and Quantum Optics and
Department of Physics, Fuzhou University, Fuzhou, Fujian 350116, China}

\affiliation{School of Physics and Astronomy, University of Nottingham, Nottingham
NG7 2RD, United Kingdom and Centre for the Mathematics and Theoretical
Physics of Quantum Non-equilibrium Systems, University of Nottingham,
Nottingham NG7 2RD, United Kingdom}

\author{Shi-Biao Zheng}

\affiliation{Fujian Key Laboratory of Quantum Information and Quantum Optics and
Department of Physics, Fuzhou University, Fuzhou, Fujian 350116, China}

\begin{abstract}
Preparation of entangled steady states via dissipation and pumping
in Rydberg atoms has been recently found to be useful for quantum
information processing. The driven-dissipative dynamics is closely
related to the natural linewidth of the Rydberg states and can be
usually modulated by engineering the thermal reservior. Instead of
modifying the effectively radiative decay, we propose an alternatively
optimized scheme, which combines the resonant Rydberg antiblockade
excitation and the Lyapunov control of the ground states to speed
up the prepration of the singlet state for two interacting Rydberg
atoms. The acceleration process strongly depends on the initial state
of the system with respect to the initial coherence between the singlet
state and decoherence-sensitive bright state. We study the optimal
parameter regime for fast entanglement preparation and the robustness
of the fidelity against random noises. The numerical results show
that a fidelity above 0.99 can be achieved around 0.4 ms with the
current experimental parameters. The scheme may be generalized for
preparation of more complicate multi-atom entangled states.
\end{abstract}
\maketitle

\section{Introduction}

Rydberg atoms are considered the ideal architecture for quantum information
processing since it provides strongly interatomic interaction on demand
and keeps a radiative lifetime as long as tens of microseconds allowing
for laser addressing \cite{GallagherTF_1994,Low_PB2012,Saffman_RMP2010,Vogt_PRL2006,Tong_PRL2004,Dudin_Science2012}.
Despite continuous experimental progress in enhancing the Rydberg-ground
coupling, it has been demonstrated that the spontaneous emission of
the high-lying excited states still causes non-negligible detrimental
effects in preparation of atomic entangled states \cite{Plenio_PRA1999,Cabrillo_PRA1999,Schneider_PRA2002,Braun_PRL2002,Jakobczyk_PRA2002,ShaoXQ_a_PRA2017,Su_PRA2015},
construction of quantum logic gate \cite{Jaksch_PRL2000-1,Su_PRA2016,Wu_PRA2017-1,Wu_PRA2010-1},
and engineering of many-body quantum dynamics \cite{Dudin_NP2012}
via Rydberg mediated interactions. Instead of protecting the Rydberg
open system against atomic decay induced decoherence, there have been
several studies \cite{CarrAW_PRL2013,Bhaktavatsala RaoDD_PRL2013,ChenX_PRA2017,ShaoXQ_PRA2014,LeeS_NJP2015,ShaoXQ_b_PRA2017,ChenX_PRA2018},
which proposed to use dissipation as a resource by Rydberg pumping.
Among these proposals, Rydberg states are excited by using a single-photon
excitation or a two-photon excitation process, and the general tasks
of quantum information processing can be realized generally in a long
time limit. Because of that, more recent interests have centered around
speeding up the pumping-dissipative dynamics toward the desired steady
state. Some first attempts have been proposed by engineering the radiative
decay with an artificial reservoir \cite{Bhaktavatsala RaoDD_PRA2014}.

Optimal control techniques can usually provide efficient and operational
algorithms for dynamical control of quantum systems \cite{HouSC_PRA2012}.
Independently of the type of quantum architecture, the general paradigm
for quantum control is guiding the quantum dynamics to attain the
desired state by engineering the system's Hamiltonian \cite{WangX_PRL2011}.
One of the useful control approaches is the Lyapunov-based control
(see \cite{HouSC_PRA2012,WangX_IEEE2010,KuangS_Automatica2008,KuangS_AAS2010,MirrahimiM_Automatica2005,GrivopoulosS_IEEE2003,VettoriP_LAIA2002,ShiZC_PRA2015,WangX_PRA2009,CuiW_PRA2013,WangX_PRA2010,YiXX_PRA2009,WangW_PRA2010}
and the references herein), which consists of using Lyapunov functions
to generate trajectories and open-loop steering control, and has the
advantage of being simple to handle for rigorous analysis. Successful
applications of the Lyapunov control have been found in control of
single-particle systems in decoherence-free subspace \cite{YiXX_PRA2009,WangW_PRA2010},
preparation of few-atom entangled states in the context of cavity
quantum electrodynamics \cite{WangX_PRA2009,CuiW_PRA2013,RanD_PRA2017,ChenYH_PRA2017,ChenYH_PRA2018},
state transfer along the spin chains \cite{ShiZC_PRA2015,WangX_PRA2010},
as well as dynamical oscillation of macroscopic objects \cite{HouSC_PRA2012}.
In general, the dynamical evolution can be accelerated to resist the
system's decoherence induced by the photon leakage and photon scattering. 

In this paper, we propose a dissipation-based scheme to prepare the
two-atom singlet state by modifying the Rydberg antiblockade condition
and the coherent unitary dynamics via Lyapunov control. We first compensate
Rydberg interactions induced level shift by the two-photon detuning
of an optical excitation laser, the Stark shifts caused by which further
matches the microwave driving frequency. By making use of the atomic
spontaneous emission as a resource and adding target-state-tracking
fields to coherently control the driven-dissipative process, we finally
generate a unique steady singlet state with high fidelity. Compared
to the previous studies, we have further examined the effect of the
Stark shifts induced by the dispersive coupling between the ground
state and the Rydberg state, which is found to be important for the
driven-dissipative dynamics. By canceling out the Stark shifts with
a detuned microwave field or an auxiliary dispersive laser, the level
configuration of the two-atom system can be effectively regarded as
a five-level resonant interaction system. Therefore, it is not difficult
to exactly solve the coherent evolution dynamics, which provides the
hints for optimal control, e.g., by appropriately selecting the optimal
frequency and driving strength of the microwave field, the converging
to the target state can be much faster than that with other choices.
On the other hand, the driven-dissipative dynamics towards the decoherence-free
state is further accelerated by using the method of Lyapunov control,
which leads to a fidelity of higher than 0.99 around 0.4 ms. In general,
our scheme gives an example for fast dissipative preparation of singlet
state for two Rydberg atoms via the Lyapunov control, and more importantly,
it would be helpful for understanding the details of the Rydberg-ground
interaction dynamics under the drivings of an optical laser and a
microwave field, which is significant for understanding the Rydberg-interaction-mediated
many-body dynamics.

The paper is organized as follows: In section 2, we derive the effective
Hamiltonian for the atoms suffering from antiblockade Rydberg interactions
and introduce the basic scheme for dissipative preparation of two-atom
singlet state. In section 3, the Lyapunov control method and its application
in speeding up the dissipative state preparation is discussed in detail.
The improved scheme under the Lyapunov control is numerically studied
in section 4, where we consider the effect of the system initial state,
atomic spontaneous emission, and the control parameters on the converging
process. In section 5, the experimental feasibility and the robustness
against the random fluctuation in parameters are further discussed.
Finally, we summarize our results in section 6.

\section{Rydberg antiblockade and effective model}

\begin{figure}
\includegraphics[width=1\columnwidth]{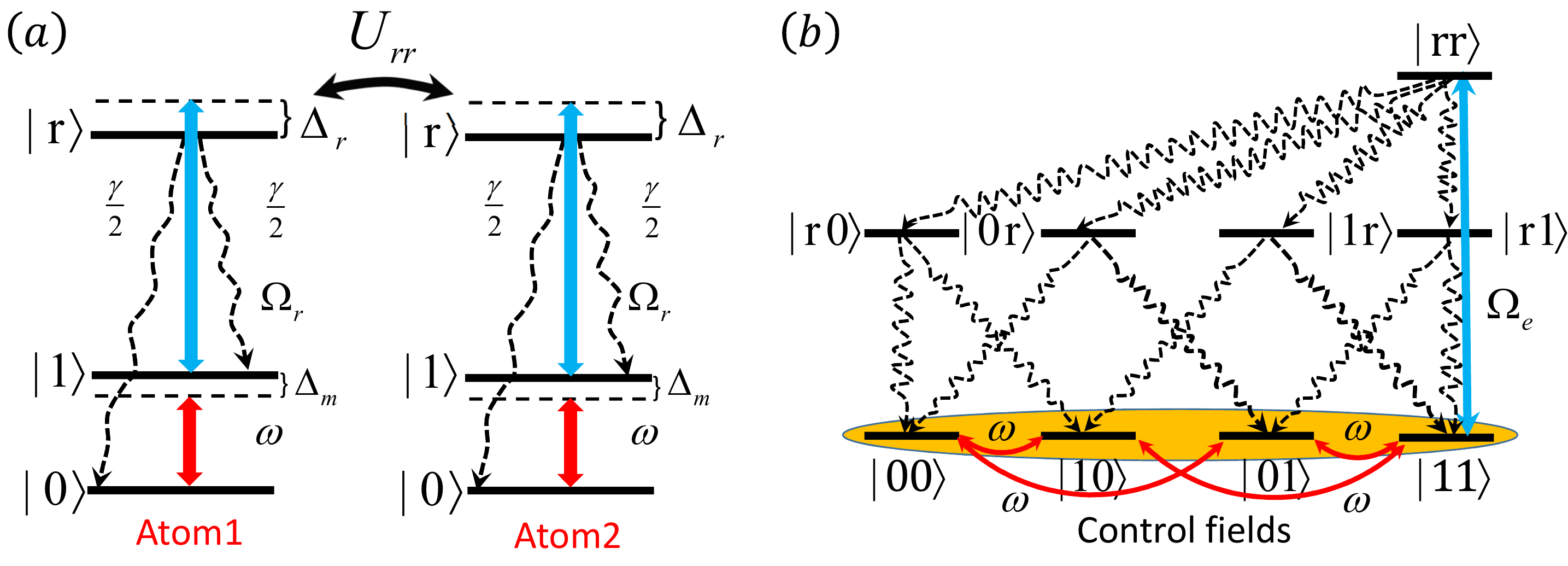}\caption{\label{Fig_atom_model}(Color online) (a) Diagram illustration of
the atomic-level configuration. The Rydberg state $|r\rangle$ is
excited from the ground state $|1\rangle$ by an optical laser with
Rabi frequency $\Omega_{r}$ and detuning $\Delta_{r}$. The ground
state transition $|0\rangle\leftrightarrow|1\rangle$ is driven by
a microwave field with Rabi frequency $\Omega_{m}$ and detuning $\Delta_{m}$.
Double excitation of the Rydberg state $|r\rangle$ is shifted by
$U_{rr}$ due to the interatomic interaction. Moreover, we assume
the Rydberg state spontaneously decays into the two ground states
at the same rate $\gamma/2$. (b) Effective level configuration for
the two-atom system and atomic transitions among the collective states.
The coherent evolution is restrained in the collective-state subspace
spanned by $|00\rangle$, $|01\rangle$, $|10\rangle$, $|11\rangle$,
and $|rr\rangle$, see main text for the detail.}
\end{figure}

We consider a system consisting of two identical Rydberg atoms, each
of them has two ground states $|0\rangle$, $|1\rangle$ and one Rydberg
excited state $|r\rangle$, as shown in Fig.\ref{Fig_atom_model}(a).
The transition $|1\rangle\leftrightarrow|r\rangle$ is driven by a
classical optical laser with Rabi frequency $\Omega_{r}$ and detuning
$\Delta_{r}$, while the transition $|0\rangle\leftrightarrow|1\rangle$
is driven by a microwave field (or alternatively by a two-photon Raman
transition) with Rabi frequency $\Omega_{m}$ and detuning $\Delta_{m}$.
A dipole-dipole interaction between the atoms arises from the simultaneous
excitation of two atoms to the Rydberg state, and the strength $U_{rr}$
of the Rydberg-mediated interaction depends on the interatomic distance
and the principal quantum number of the Rydberg state. The Rydberg
population spontaneously decays onto the ground states $|0\rangle$
and $|1\rangle$ with the same rate $\gamma/2$ due to the finite
radiative lifetime. The Hamiltonian for the system, in the interaction
picture, is given by ($\hbar=1$) 

\begin{equation}
H=H_{l}+H_{m},\label{eq:the_original_Hamiltonian}
\end{equation}
with
\[
H_{l}=\frac{1}{2}\sum_{j=1,2}(-\Delta_{r}|r\rangle_{j}\langle r|+\Omega_{r}|r\rangle_{j}\langle1|+H.c.)+U_{rr}|rr\rangle\langle rr|,
\]
\[
H_{m}=\frac{1}{2}\sum_{j=1,2}(\Delta_{m}|0\rangle_{j}\langle0|+\Omega_{m}|1\rangle_{j}\langle0|)+H.c..
\]
While consider the decoherence effect induced by atomic spontaneous
emissions, the dissipative dynamics of the system is then governed
by the Lindblad-Markovian master equation, i.e.
\begin{eqnarray}
\stackrel{.}{\rho} & = & -i[H,\rho]+\mathcal{L}[\rho],
\end{eqnarray}
with
\[
\mathcal{L}[\rho]=\sum_{j=1,2}\sum_{k=0,1}L_{j,k}\rho L_{j,k}^{\dagger}-\frac{1}{2}(L_{j,k}^{\dagger}L_{j,k}\rho+\rho L_{j,k}^{\dagger}L_{j,k}),
\]
where $\rho$ is the density operator of the system and $L_{j,k}=\sqrt{\gamma/2}|k\rangle_{j}\langle r|$
($k=0,1$) are Lindblad dissipation operators. 

In the regime of the Rydberg antiblockade (namely, $U_{rr}=2\Delta_{r}$)
and in the limit of the large detuning $\Delta_{r}\gg\Omega_{r}$,
the two atoms tend to be excited in bunching under the laser driving.
Then, we can derive an effective Hamiltonian for the system by using
the time averaging method \cite{James_CJP2007}, 

\begin{eqnarray}
H^{\prime} & = & 2\Delta_{m}|00\rangle\langle00|+\frac{\Omega_{r}^{2}}{2\Delta_{r}}(|11\rangle+|rr\rangle)(\langle11|+\langle rr|)\nonumber \\
 &  & +(\Delta_{m}+\frac{\Omega_{r}^{2}}{4\Delta_{r}})(|01\rangle\langle01|+|10\rangle\langle10|)\nonumber \\
 &  & +\frac{\Omega_{m}}{2}[(|11\rangle+|00\rangle)(\langle01|+\langle10|)+H.c.],\label{eq:without rotation_Hamiltonian}
\end{eqnarray}
where the basis states in the single excitation subspace are adiabatically
eliminated. We note that the dispersive coupling not only enables
the two-photon atomic transition to the doubly excitation state $|rr\rangle$,
but also introduces the Stark shifts $\sim\Delta_{r}^{-1}$ to the
collective ground states $|01\rangle$, $|10\rangle$, and $|11\rangle$.
Thus, the degeneracy in the ground state subspace is broken, namely,
a resonant driving with microwave fields to the individual atomic
transition $|0\rangle\leftrightarrow|1\rangle$ is not in resonance
with both the collective transitions $|01\rangle(|10\rangle)\leftrightarrow|00\rangle$
and $|01\rangle(|10\rangle)\leftrightarrow$$|11\rangle$ any more.
The coherent population transfer among the collective ground states
becomes less efficient.

To cancel out the effect of the Stark shifts, it is convenient to
introduce a finite detuning $\Delta_{m}=\Omega_{r}^{2}/4\Delta_{r}$
for the microwave driving. In this case, the Hamiltonian (\ref{eq:without rotation_Hamiltonian})
in the rotating frame with respect to $\Omega_{r}^{2}/2\Delta_{r}$
reduces to 
\begin{equation}
H_{e}=\frac{\Omega_{m}}{\sqrt{2}}(|11\rangle+|00\rangle)\langle B|+\frac{\Omega_{e}}{2}|11\rangle\langle rr|+H.c.,\label{eq:effective_H}
\end{equation}
where $\Omega_{e}=\Omega_{r}^{2}/\Delta_{r}$ and $|B\rangle=(|01\rangle+|10\rangle)/\sqrt{2}$.
Alternatively, we can utilize a laser field, dispersively coupling
$|1\rangle$ to the an auxiliary state and introduce a energy correction
$-\Omega_{r}^{2}/\Delta_{r}$ to the state $|1\rangle$, which leads
to the effective Hamiltonian in Eq. (\ref{eq:effective_H}) as well.
For convenience, we further define
\begin{eqnarray}
a & = & \sqrt{\Omega_{m}^{2}+\left(\frac{\text{\ensuremath{\Omega_{e}}}}{2}\right)^{2}},\nonumber \\
b^{2} & = & \sqrt{\Omega_{m}^{4}+\left(\frac{\text{\ensuremath{\Omega_{e}}}}{2}\right)^{4}},\label{eq:abc_expr}\\
c & = & \sqrt{\Omega_{m}^{2}+\frac{\Omega_{e}^{2}}{2}},\nonumber 
\end{eqnarray}
then the eigenstates of the Hamiltonian (\ref{eq:effective_H}) without
normalization are given by 
\begin{eqnarray*}
|\phi_{1}\rangle & \sim & |10\rangle-|01\rangle
\end{eqnarray*}
\begin{eqnarray*}
|\phi_{2}\rangle & \sim & [-2c^{2}\sqrt{a^{2}+b^{2}}+2(a^{2}+b^{2})^{\frac{3}{2}}]|00\rangle\\
 & + & [2\Omega_{m}(a^{2}+b^{2})-\Omega_{m}\Omega_{e}^{2}]|B\rangle\\
 & + & 2\sqrt{a^{2}+b^{2}}\Omega_{m}^{2}|11\rangle+\sqrt{2}\Omega_{m}^{2}\Omega_{e}|rr\rangle
\end{eqnarray*}

\begin{eqnarray*}
|\phi_{3}\rangle & \sim & [-2c^{2}\sqrt{a^{2}-b^{2}}+2(a^{2}-b^{2})^{\frac{3}{2}}]|00\rangle\\
 & + & [2\Omega_{m}(a^{2}-b^{2})-\Omega_{m}\Omega_{e}^{2}]|B\rangle\\
 & + & 2\sqrt{a^{2}-b^{2}}\Omega_{m}^{2}|11\rangle+\sqrt{2}\Omega_{m}^{2}\Omega_{e}|rr\rangle
\end{eqnarray*}

\begin{eqnarray*}
|\phi_{4}\rangle & \sim & [2c^{2}\sqrt{a^{2}+b^{2}}-2(a^{2}+b^{2})^{\frac{3}{2}}]|00\rangle\\
 & + & [2\Omega_{m}(a^{2}+b^{2})-\Omega_{m}\Omega_{e}^{2}]|B\rangle\\
 & - & 2\sqrt{a^{2}+b^{2}}\Omega_{m}^{2}|11\rangle+\sqrt{2}\Omega_{m}^{2}\Omega_{e}|rr\rangle
\end{eqnarray*}

\begin{eqnarray*}
|\phi_{5}\rangle & = & [2c^{2}\sqrt{a^{2}-b^{2}}-2(a^{2}-b^{2})^{\frac{3}{2}}]|00\rangle\rangle\\
 & + & [2\Omega_{m}(a^{2}-b^{2})-\Omega_{m}\Omega_{e}^{2}]|B\rangle\\
 & - & 2\sqrt{a^{2}-b^{2}}\Omega_{m}^{2}|11\rangle+\sqrt{2}\Omega_{m}^{2}\Omega_{e}|rr\rangle
\end{eqnarray*}
which are associated with the eigenvalues
\begin{equation}
E_{1}=0,\textrm{ }E_{2,3}=\frac{1}{\sqrt{2}}\sqrt{a^{2}\pm b^{2}},\textrm{ }E_{4,5}=-\frac{1}{\sqrt{2}}\sqrt{a^{2}\pm b^{2}},\label{eq:the_eigenvalue}
\end{equation}
respectively. Thus, the instantaneous state of the system under the
coherent evolution can be written as the superposition of the five
eigenstates, namely 
\begin{equation}
|\psi(t)\rangle=\sum_{k=1}^{5}C_{k}(t)|\phi_{k}\rangle.\label{eq:Coh_evol}
\end{equation}

\begin{figure}
\includegraphics[width=1\columnwidth]{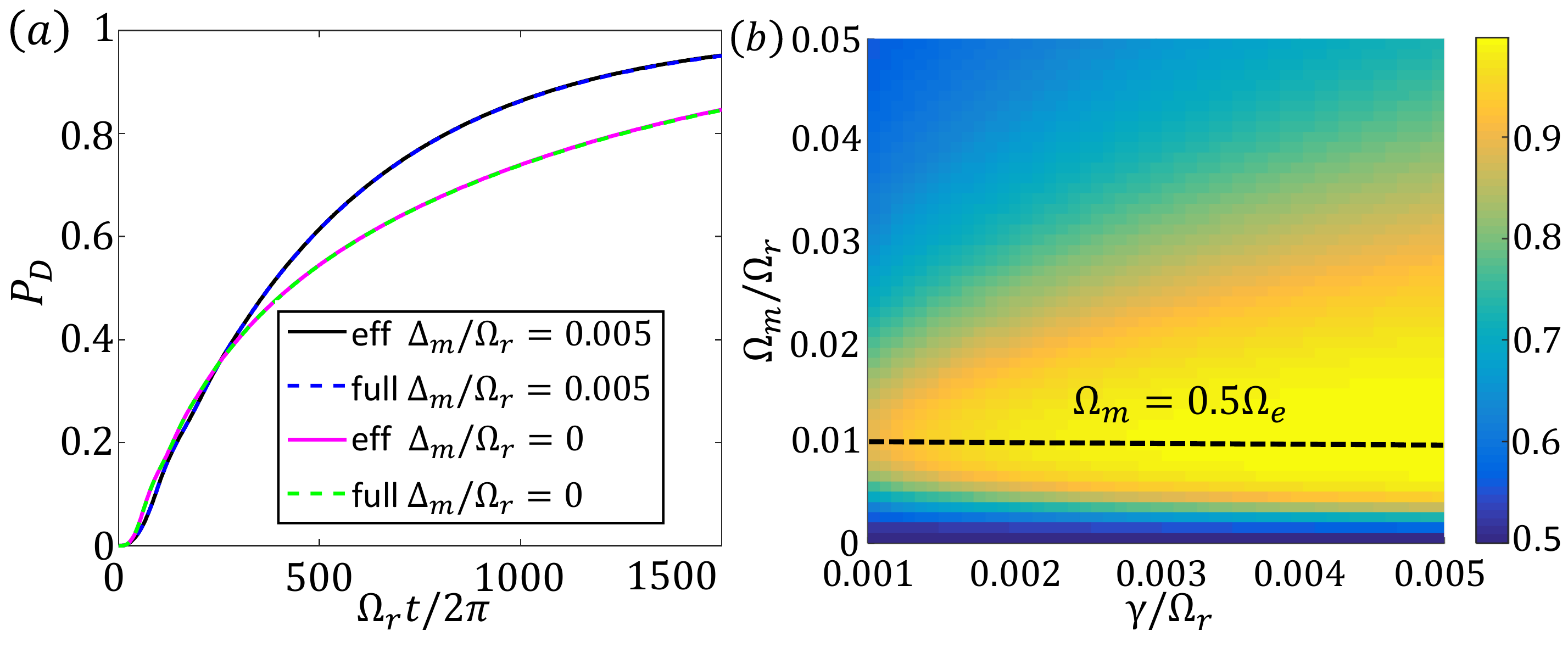}\caption{\label{Fig_Fide_nhc}(Color online) (a) Population of the singlet
state $P_{D}=\textrm{Tr}[\rho(t)|D\rangle\langle D|]$ as a function
of rescaled time with the system initially being in $(|00\rangle+|01\rangle+|10\rangle+|11\rangle)/\sqrt{2}$.
The time evolution governed by the effective Hamiltonian (\ref{eq:effective_H})
agrees well with that dominated by the originally full Hamiltonian
(\ref{eq:the_original_Hamiltonian}). The microwave driving with a
slight detuning gives rise to a faster convergence to the target state
compared with that under resonant driving. (b) $P_{D}$ versus the
Rabi frequency $\Omega_{m}$ of the microwave driving and the spontaneous
emission rate $\gamma$ of the Rydberg state for $\Omega_{r}t/2\pi=1500$.
The optimal driving scheme for the microwave field is found for $\Omega_{m}=0.5\Omega_{e}$.
Other parameters are $(\Delta_{r},\gamma,\Omega_{m})/\Omega_{r}=(50,0.002,0.01)$
in (a) and $(\Delta_{r},\Delta_{m})/\Omega_{r}=(50,0.005)$ in (b)
in units of $\Omega_{r}=1$.}
\end{figure}

It has been clearly shown that there exists a unique dark state $|D\rangle=\frac{1}{\sqrt{2}}(|10\rangle-|01\rangle)$,
which is also referred to as the two-atom singlet state, corresponding
to the null eigenvalue of the system Hamiltonian (\ref{eq:the_original_Hamiltonian}).
We now turn to the scheme for preparation of the singlet state by
considering and taking advantage of the atomic spontaneous emission. 

\begin{figure*}[t]
\includegraphics[width=0.8\textwidth]{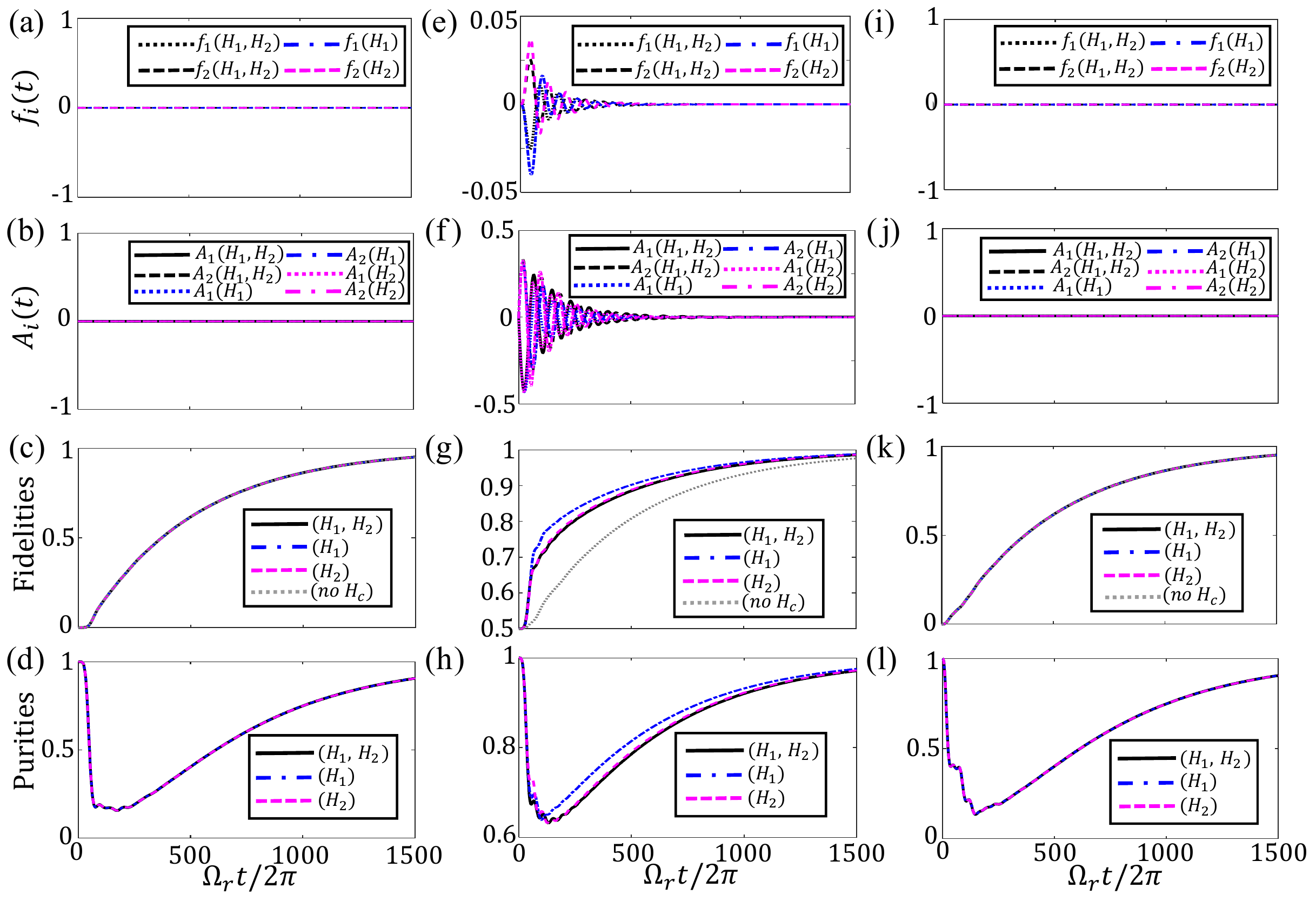}\caption{\label{fig:fAFP_vs_Time}(Color online) Control functions $f_{j}(t)$
{[}Eq.(\ref{eq:Lyp_function}){]}, matrix elements $A_{j}$ ($j=1,2$),
the fidelity and purity of the target state $|D\rangle$ versus dimensionless
time for the atoms initially being in the state $|00\rangle$ {[}(a)-(d){]},
$|10\rangle$ {[}(e)-(h){]}, $|11\rangle$ {[}(i)-(l){]}. The system
is subject to three different control manners: (1) $H_{1}\protect\neq0$,
$H_{2}=0$, (2) $H_{1}=0$, $H_{2}\protect\neq0$, and (3) $H_{1}\protect\neq0$,
$H_{2}\protect\neq0$, respectively. Other parameters are $(\Delta_{r},\gamma,\Omega_{m},\Delta_{m},\lambda_{1},\lambda_{2})/\Omega_{r}=(50,0.002,0.01,0.005,0.08,0.08)$
in units of $\Omega_{r}=1$.}
\end{figure*}

The schematic diagram for preparing the singlet state is shown in
Fig. \ref{Fig_atom_model}(b), where the collective level configuration
consists of four ground states $|00\rangle$, $|01\rangle$, $|10\rangle$,
$|11\rangle$, four single-excitation states $|0r\rangle$, $|1r\rangle$,
$|r0\rangle$, $|r1\rangle$, and a doubly excitation state $|rr\rangle$.
Without the optical excitation of the Rydberg state, the two-atom
system will stay in the subspace spanned by \{$|00\rangle$, $|B\rangle$,
$|11\rangle$\} under the microwave driving $H_{m}$. While the optical
laser is applied, only when the system populates the state $|11\rangle$
will it be excited to the doubly Rydberg state $|rr\rangle$ (i.e.
the Rydberg pumping process), which then decays onto the single-excitation
subspace followed by re-occupation of the ground states. Based on
the effective Hamiltonian (\ref{eq:effective_H}), the driven-dissipative
dynamics can now be described by
\begin{eqnarray}
\overset{\cdot}{\rho} & = & -i\left[H_{e},\rho\right]+\mathcal{\widetilde{L}}[\rho]\label{eq:Master_Eq}
\end{eqnarray}
with $\widetilde{L}_{j,k}=\sqrt{\gamma/2}|k\rangle_{j}\langle r|\otimes\mathcal{I}_{j^{\prime}\neq j}$
($j$, $j^{\prime}$ $=1$ or 2). It turns out that the singlet state
$|D\rangle$ is a decoherence-free state, which is decoupled from
any of the driving fields and is the unique steady state solution
of the master equation (\ref{eq:Master_Eq}), see the example with
the initial state $(|00\rangle+|01\rangle+|10\rangle+|11\rangle)/\sqrt{2}$
shown in Fig. \ref{Fig_Fide_nhc}(a). Mathematically, this can be
verified simply by setting $\dot{\rho}=0$ and checking $i\left[H_{e},\rho_{ss}\right]=\mathcal{\widetilde{L}}[\rho_{ss}]$
with $\rho_{ss}$=$|D\rangle\langle D|$. In addition, with the choice
of the detuning $\text{\ensuremath{\Delta}}_{m}=\Omega_{r}^{2}/4\Delta_{r}$
and the optimal driving strength $\Omega_{m}=\Omega_{e}/2$ {[}which
corresponds to $a=\sqrt{2}\Omega_{m}$, $b^{2}=\sqrt{2}\Omega_{m}^{2}$,
and $c=\sqrt{3}\Omega_{m}$, see Eq. (\ref{eq:abc_expr}){]}, the
system converges to the steady state faster than that with resonant
microwave driving. However, this regular driven-dissipative scheme
still takes a time as long as $\Omega_{r}t/2\pi\sim1000$ to achieve
a population of the singlet state $P_{D}=\textrm{Tr}[\rho(t)\rho_{ss}]$
larger than 0.9.

\section{Lyapunov control with microwave fields}

The converging process to the steady state can be further accelerated
by following the algorithm of the Lyapunov control, which is a method
of local optimal control with numerous variants and possesses the
advantages of robustness and stability \cite{SugawaraM_JCP2003}.
To realize an additional control of the system, we first introduce
the control Hamiltonian
\begin{equation}
H_{c}=\sum_{j=1,2}f_{j}(t)H_{j},
\end{equation}
where $H_{j}=\lambda_{j}|0\rangle_{j}\langle1|+h.c.$ and the time
varying control functions $f_{j}(t)$ are new degree of freedom that
allow us to design the system evolution towards the desired state
$|D\rangle$. The coherent dynamics of the system is now governed
by the quantum Liouville equation $\stackrel{.}{\rho}=-i[H_{e}+H_{c},\rho]$.
According to the Lyapunov control theory \cite{HouSC_PRA2012,WangX_IEEE2010,KuangS_Automatica2008,KuangS_AAS2010,MirrahimiM_Automatica2005,GrivopoulosS_IEEE2003,VettoriP_LAIA2002,ShiZC_PRA2015,WangX_PRA2009,CuiW_PRA2013,WangX_PRA2010,YiXX_PRA2009,WangW_PRA2010},
the control fields $f_{j}(t)$ can be given via pulse shaping of the
laser intensities corresponding to a pre-defined Lyapunov function
$\xi(t)$, which fulfills the sufficient conditions $\xi(t)\geq0$
and $\partial_{t}\xi(t)\leq0$.

In order to describe the converging efficiency of the system under
Lyapunov control, we then introduce the time-dependent 'distance'
of the system state away from the target state $S(t)\equiv1-F(t)\geq0$,
with $F(t)=\langle D|\rho(t)|D\rangle$ refers to the instantaneous
fidelity of the target state $|D\rangle$. Moreover, the time-derivative
of the 'distance' $V(t)=\partial_{t}S(t)$ is used for characterizing
the instantaneous evolution speed of the driven-dissipative process,
leading to \cite{HouSC_PRA2012,WangX_IEEE2010,KuangS_Automatica2008,KuangS_AAS2010,MirrahimiM_Automatica2005,GrivopoulosS_IEEE2003,VettoriP_LAIA2002,ShiZC_PRA2015,WangX_PRA2009,CuiW_PRA2013,WangX_PRA2010,YiXX_PRA2009,WangW_PRA2010}
\begin{eqnarray}
V(t) & = & -\langle D|\stackrel{\cdotp}{\rho}(t)|D\rangle=V_{a}(t)+V_{b}(t),\label{eq:the_distace_function}
\end{eqnarray}
with
\[
V_{a}(t)=-\sum_{j=1,2}f_{j}(t)\langle D|[-iH_{j},\rho(t)]|D\rangle,
\]
\[
V_{b}(t)=-\sum_{j,k}\langle D|\widetilde{L}_{j,k}\rho\widetilde{L}_{j,k}^{\dagger}|D\rangle,
\]
where we have considered the fact that the decoherence free state
$|D\rangle$ is a stationary state satisfying $H_{e}|D\rangle=0$
and $\widetilde{L}_{j,k}|D\rangle=0$. Our main task in the next step
is to design a dynamical control that ensures that $S(t)$ is monotonically
decreasing until the end of system evolution, which requires $V(t)\leq0$.
Note that the second term in Eq. (\ref{eq:the_distace_function})
can be diagonalized in terms of the effective orthonormal excited
states \{$|E_{m}\rangle$\}, giving rise to $V_{b}(t)=-\sum_{m}\gamma_{m}\langle E_{m}|\rho(t)|E_{m}\rangle\leq0$,
with $\gamma_{m}$ being the effective decay rates. Then, if we further
set
\begin{eqnarray}
f_{j}(t) & = & -i\langle D|[H_{j},\rho]|D\rangle\nonumber \\
 & = & \sqrt{2}\lambda_{j}\textrm{Im}(\langle11|\rho|D\rangle+\langle D|\rho|00\rangle),\label{eq:Lyp_function}
\end{eqnarray}
it is easy to verify $V_{b}(t)\leq0$ and $V(t)\leq0$ with equality
only true for the system being initially in $|D\rangle$. With these
choices, $S(t)$ becomes exactly the Lyapunov function we are seeking
for. Since the control fields $f_{j}(t)$ are target-state-dependent
functions and are determined by the elements $A_{1}(t)=\textrm{Im}(\langle11|\rho|D\rangle),$
$A_{2}(t)=\textrm{Im}(\langle D|\rho|00\rangle)$ of the density matrix,
they vanish at the end of dynamical evolution leaving the system in
the steady state $|D\rangle$. This time-varying optimization method
is also known as trajectory tracking control \cite{HouSC_PRA2012,WangX_IEEE2010,KuangS_Automatica2008,KuangS_AAS2010,MirrahimiM_Automatica2005,GrivopoulosS_IEEE2003,VettoriP_LAIA2002,ShiZC_PRA2015,WangX_PRA2009,CuiW_PRA2013,WangX_PRA2010,YiXX_PRA2009,WangW_PRA2010}.

\section{Numerical results and discussions}

The driven-dissipative dynamics of the system subjected to the Lyapunov
control is now dominated by 
\begin{eqnarray}
\overset{\cdot}{\rho} & = & -i\left[H_{e}+H_{c},\rho\right]+\mathcal{\widetilde{L}}[\rho],\label{eq:Master_LYP}
\end{eqnarray}
based on which, we have shown in Fig. \ref{fig:fAFP_vs_Time} the
time-dependent Lyapunov control functions $f_{j}(t)$, the related
density matrix elements $A_{j}$ ($j=1,2$), the fidelity $F(t)$
and the purity $P(t)=\text{Tr}(\rho^{2})$ of the system with the
atoms initially being in the state $|00\rangle$, $|10\rangle$, and
$|11\rangle$, respectively. 

We first assume that the system is initially in the state $|\psi(0)\rangle=|00\rangle$.
In this case, the fidelity of the singlet state goes through a transient
state with $F\approx0$ and then surpasses $0.9$ at the moment around
$\Omega_{r}t/2\pi\sim1150$, as shown in Fig. \ref{fig:fAFP_vs_Time}(a)-\ref{fig:fAFP_vs_Time}(d).
The driven-dissipative dynamics is not accelerated simply because
the Lyapunov control cannot be triggered with the matrix elements
$\sim A_{1,2}(t)$ of the density operator being zero at all time,
i.e. $A_{1,2}(0)\sim\langle D|00\rangle=0$. For the system initially
in the state $|\psi(0)\rangle=|11\rangle$, the evolutional dynamics
does not exhibit a transient behavior due to the immediate excitation
of the Rydberg doubly excitation state. But Similarly, the converging
process can not be speeded up due to $A_{1,2}(0)\sim\langle D|11\rangle=0$,
see Fig. \ref{fig:fAFP_vs_Time}(i)-\ref{fig:fAFP_vs_Time}(l).

While the system is initially in the state $|\psi(0)\rangle=|10\rangle$
or $|\psi(0)\rangle=|01\rangle$, the Lyapunov control takes effect
and the fidelity of larger than 0.9 for the target state can be achieved
in a greatly reduced time $\Omega_{r}t/2\pi\sim600$, as shown in
Fig. \ref{fig:fAFP_vs_Time}(e)-\ref{fig:fAFP_vs_Time}(h). Essentially,
this is due to the fact that $|10\rangle$ ($|01\rangle$) consists
in the coherence between the bright state $|B\rangle$ and the singlet
state $|D\rangle$, which is nevertheless non-existent in $|00\rangle$
and $|11\rangle$. To see the insight, we rewrite the initial state
$|10\rangle$ as
\begin{equation}
|\psi(0)\rangle=\frac{1}{2}(|B\rangle\langle B|+|D\rangle\langle D|+|D\rangle\langle B|+|B\rangle\langle D|).
\end{equation}
Since the bright state $|B\rangle$ under the dominance of the system
Hamiltonian (\ref{eq:effective_H}) coherently couples to both $|00\rangle$
and $|11\rangle$, which immediately transforms the third and fourth
terms into the nonvanishing matrix elements $\langle11|\rho|D\rangle$
and $\langle D|\rho|00\rangle$, activating the self-adaptive control
functions {[}see Eq. (\ref{eq:Lyp_function}){]}. As the dissipative
dynamics evolves and the coherence relaxation between the singlet
and the dark state increases, the effect of the Lyapunov control gradually
vanishes and finally the conventional driven-dissipative dynamics
dominates. We note that the scheme with simply the single-atom Lyapunov
control (i.e. $H_{c}=f_{1}(t)H_{1}$ for $|\psi(0)\rangle=|10\rangle$
or $H_{c}=f_{2}(t)H_{2}$ for $|\psi(0)\rangle=|01\rangle$) can even
achieve a better converging effect, which arises from the symmetry
breaking of the coherent evolution among $|00\rangle$, $|B\rangle$
and $|11\rangle$. 

\begin{figure}
\includegraphics[width=1\columnwidth]{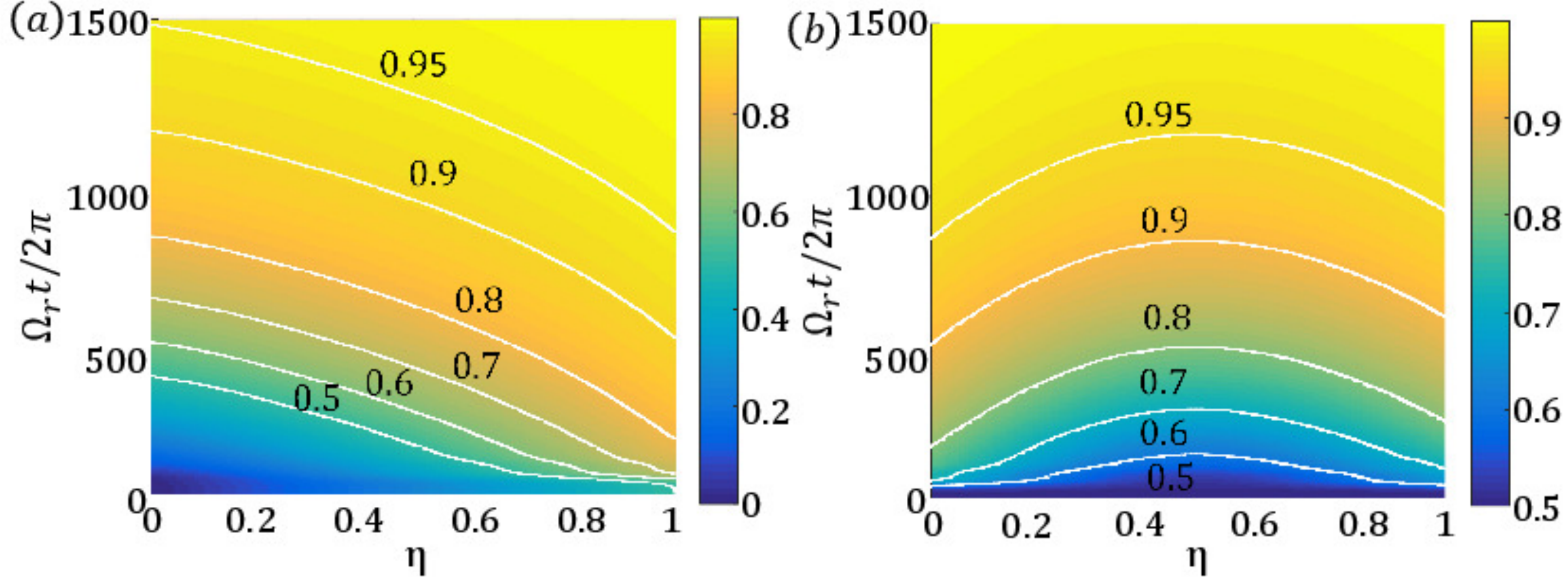}\caption{\label{fig:Fide_MixIniS}(Color online) Time-dependent fidelity of
the singlet state $|D\rangle$ under the Lyapunov control $H_{1}$
for mixed initial states (a) $(1-\eta)|00\rangle\langle00|+\eta|10\rangle\langle10|$
and (b) $(1-\eta)|10\rangle\langle10|+\eta|01\rangle\langle01|$,
respectively. Parameters are $(\Delta_{r},\gamma,\Omega_{m},\Delta_{m},\lambda_{1},\lambda_{2})/\Omega_{r}=(50,0.002,0.01,0.005,0.08,0)$
and $\Omega_{r}=1$.}
\end{figure}

Moreover, we look into the controlled dynamics with the initial mixed
states

\begin{eqnarray}
|\psi(0)\rangle & = & (1-\eta)|00\rangle\langle00|+\eta|10\rangle\langle10|
\end{eqnarray}
and 
\begin{eqnarray}
|\psi(0)\rangle & = & (1-\eta)|10\rangle\langle10|+\eta|01\rangle\langle01|\nonumber \\
 & = & \frac{1}{2}(|B\rangle\langle B|+|D\rangle\langle D|)+(\frac{1}{2}-\eta)(|D\rangle\langle B|\nonumber \\
 &  & +|B\rangle\langle D|),
\end{eqnarray}
respectively, see Fig. \ref{fig:Fide_MixIniS}. The former leads to
the intuitive result that the converging speed gradually increases
as the initial population of the basis state $|10\rangle$ grows.
For the latter case, we find that the slowest converging speed appears
at $\eta=0.5$, which corresponds to the initial state $\rho(0)=\rho_{BD}\equiv(|B\rangle\langle B|+|D\rangle\langle D|)/2$.
Although the population of the bright state $|B\rangle$ and the singlet
state $|D\rangle$ for the initial states $|10\rangle\langle10|$
and $\rho_{BD}$ are the same, the completely mixed state $\rho_{BD}$
does not possess any coherence between them, which inhibits the effect
of the control functions and again confirms our prediction with respect
to the speed-up conditions.

\begin{figure}
\includegraphics[width=1\columnwidth]{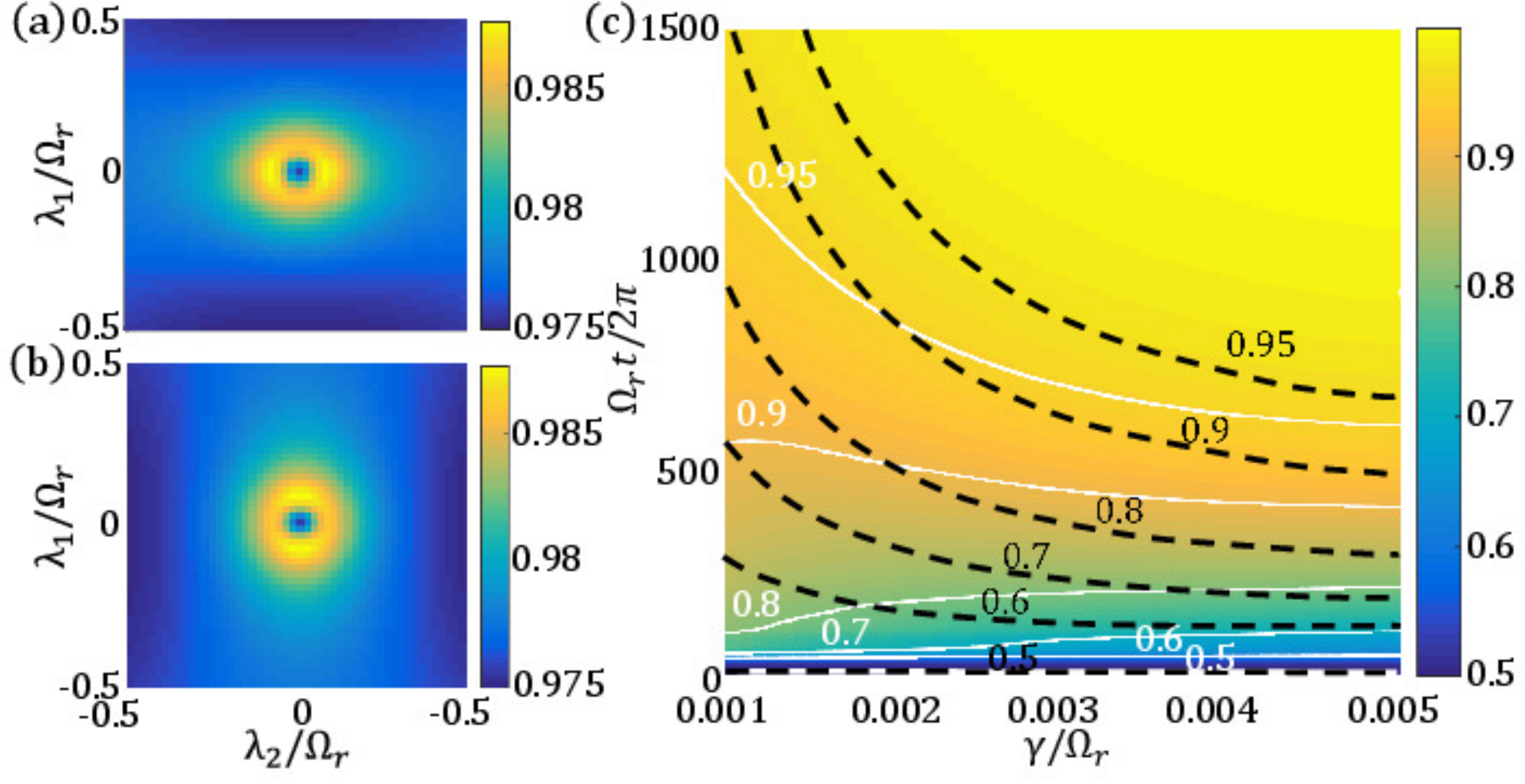}\caption{\label{fig:Opt_LC_St} (Color online) Fidelity of the singlet state at time $\Omega_{r}t/2\pi=1500$
as functions of the control parameters $\lambda_{1}$ and $\lambda_{2}$
with the system initially being in $|01\rangle$ {[}(a){]} and $|10\rangle$
{[}(b){]}, and as functions of atomic spontaneous emission rate $\gamma/\Omega_{r}$
and dimensionless time $\Omega_{r}t$ {[}(c){]}. In (c), the white
(solid) and black (dash) contour lines denote the fidelities with
and without Lyapunov control, respectively. Other parameters are the
same as in Fig. \ref{fig:Fide_MixIniS}.}
\end{figure}

Although the additive Lyapunov control can accelerate the preparation
of the singlet state, it does not imply that the stronger the driving
intensity of the control fields applies the faster the driven-dissipative
process converges, which is evidenced in Fig. \ref{fig:Opt_LC_St}(a)-\ref{fig:Opt_LC_St}(b).
For the initial states $|01\rangle$ ($|10\rangle$), the optimized
driving strengths appear at $\lambda_{1}=0$, $\lambda_{2}=0.8$ ($\lambda_{1}=0.8$,
$\lambda_{2}=0$), which are asymmetric in driving intensities of
the two microwave field. Finally, we point out that the acceleration
effect strongly depends on the decay rate $\gamma\sim n^{2}$ (or
the principal quantum number $n$) of the Rydberg state, as shown
in Fig. \ref{fig:Opt_LC_St}(c). For a slow dissipation process, the
Lyapunov control can significantly increase the converging efficiency,
e.g., for $\gamma/\Omega_{r}=0.001$, the fidelity of higher 0.9 can
be achieved at the time about $1/3$ of that without Lyapunov control.
In contrast, only a modest extent of acceleration can be reached for
a fast decay for $\gamma/\Omega_{r}\sim0.005$. 

\section{Experimental feasibility and Influence of the stochastic parameter
fluctuations }

In the context of experimental feasibility, the schematic energy-level
diagram can be encoded by the clock states $|0\rangle=|6S_{1/2},F=3\rangle$,
$|1\rangle=|6S_{1/2},F=4\rangle$ and the Rydberg state $|r\rangle=|64P_{3/2}\rangle$
in the $^{133}$Cs atoms \cite{JauYY_Nature_physics2016}. The transition
from the ground state $|1\rangle$ to the Rydberg state $|r\rangle$
is driven by a 390-nm single-photon excitation laser with the Rabi
frequency $\Omega_{r}/2\pi\sim4$ MHz. The dynamical oscillation between
the two clock states $|0\rangle$ and $|1\rangle$ is controlled by
microwave fields with the coupling strength up to $1$MHz. The decay
rate ( or the life time) of the Rydberg state $|r\rangle$ is $\gamma=2\pi\times0.007$
MHz ($\sim150\text{ \ensuremath{\mu}}$s). Therefore, the parameter
regime we considered is within the reach of the state of the art Rydberg
experiments. With respect to the particular case \cite{JauYY_Nature_physics2016},
the time required to obtain a high-fidelity $(\text{\ensuremath{\sim}}99\%)$
steady-state entanglement is only about $0.4$ ms.
\begin{figure}
\includegraphics[width=1\columnwidth]{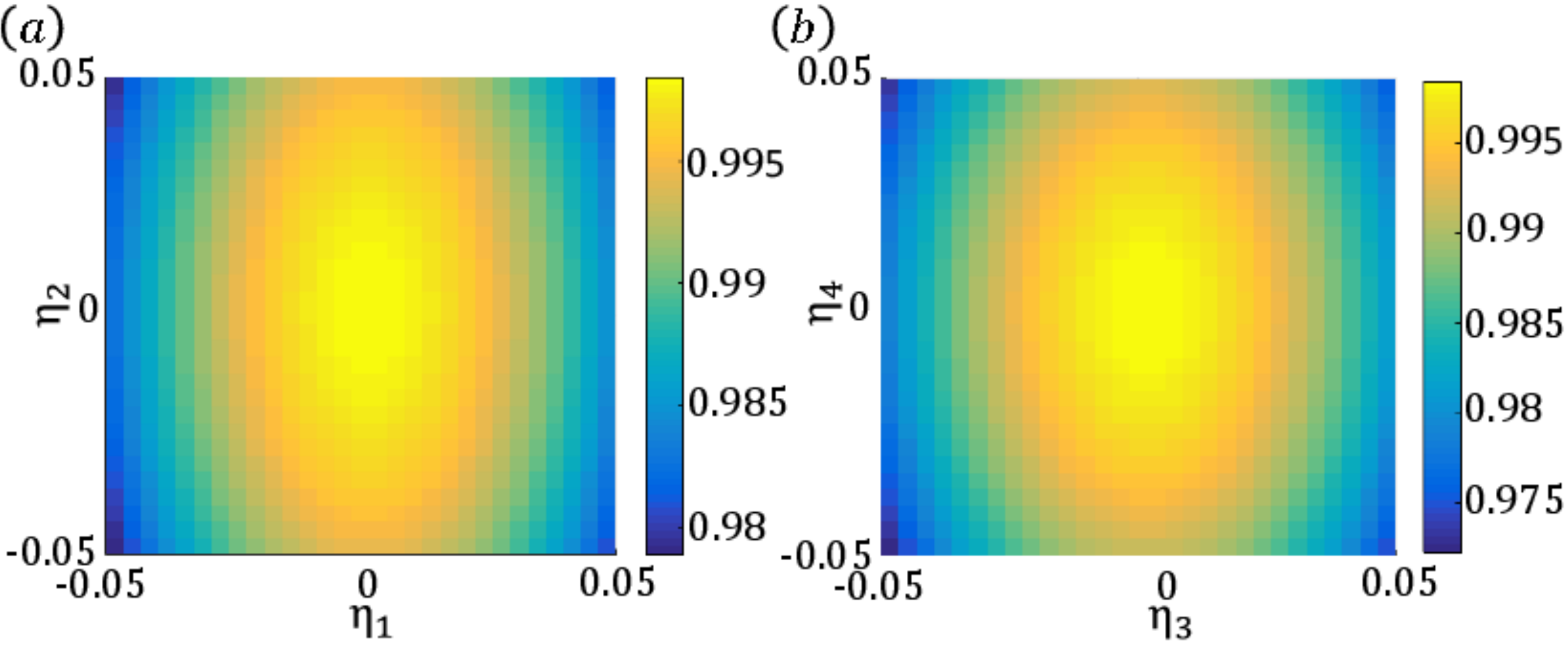}\caption{\label{Fig_Fidelity_vs_noise} (Color online) Robustness of the steady-state fidelity
for $\Omega_{r}t/2\pi=2500$ with respect to stochastic amplitude
noise in microwave driving strength ($\eta_{1}$) and frequency ($\eta_{2}$),
optical excitation Rabi frequency ($\eta_{3}$), and interatomic interaction
($\eta_{4}$). We assume the initial state of system is $|10\rangle$
and the ideal evolution dynamics is exactly the same to that in Fig.
\ref{fig:fAFP_vs_Time}(e)-(h) before adding the stochastic noise.}
\end{figure}

Furthermore, we investigate the influence of stochastic fluctuations
in parameters, such as laser intensity $\Omega_{r}$, microwave driving
strength $\Omega_{m}$ and detuning $\Delta_{m}$, as well as Rydberg-Rydberg
interactions $U_{rr}$ on the steady-state fidelity. To describe the
stochastic process, we assumed that the system Hamiltonian is now
composed of the coherent part $H_{0}$ and the amplitude-noise part
$\eta_{k}\xi(t)H_{sk}$ ($k=1,2,3,4$) with
\begin{eqnarray}
H_{s1} & = & \frac{\Omega_{m}}{2}\sum_{j=1,2}|0\rangle_{j}\langle1|+H.c.,\nonumber \\
H_{s2} & = & \frac{\Delta_{m}}{2}\sum_{j=1,2}|0\rangle_{j}\langle0|+H.c.,\nonumber \\
H_{s3} & = & \frac{\Omega_{r}}{2}\sum_{j=1,2}|1\rangle_{j}\langle r|+H.c.,\nonumber \\
H_{s4} & = & \frac{U_{rr}}{2}|rr\rangle\langle rr|+H.c.,
\end{eqnarray}
and $\xi(t)$ being the Gaussian white noise satisfying $\langle\xi(t)\rangle=0$
and $\langle\xi(t)\xi(t^{\prime})\rangle=\delta(t-t^{\prime})$. Starting
from an initial pure state $\rho_{\xi}(t=0)$, the stochastic dynamics
of the system without involving the atomic decay is governed by 
\begin{eqnarray}
\stackrel{\cdot}{\rho}_{\xi} & = & -i[H_{0},\rho_{\xi}]-i\eta[H_{sk},\xi\rho_{\xi}],\label{eq:sto_master_without dissipation}
\end{eqnarray}
which after averaging over the random trajectories becomes \cite{Ruschhaupt_NJP2014}
\begin{eqnarray}
\stackrel{\cdot}{\rho} & \simeq & -i[H_{0},\rho]-i\eta[H_{sk},\langle\xi\rho_{\xi}\rangle].
\end{eqnarray}
Using the Novikov's theorem, the average of the product of the noise
and the noise-dependent density matrix can be further given by $\langle\xi\rho_{\xi}\rangle=\frac{1}{2}\langle\frac{\delta\rho_{\xi}}{\delta\xi(t^{\prime})}\rangle|_{t=t^{\prime}}=-\frac{i\eta}{2}[H_{sk},\rho]$.
Therefore, when both the stochastic noise and the atomic dissipation
are taken into account, the evolution of the system is finally governed
by 
\begin{eqnarray}
\stackrel{\cdot}{\rho} & = & -i[H,\rho]+\mathcal{L}[\rho]+\mathbb{\mathbb{\mathcal{D}}}[\rho]\label{eq:stochastic_master_with dissipation}
\end{eqnarray}
where $\mathbb{\mathbb{\mathcal{D}}}[\rho]=-\eta_{k}^{2}[H_{sk},[H_{sk},\rho]]/2$.
Consider the single-atom Lyapunov control $H_{c}=f_{1}(t)H_{1}$ with
the control function $f_{1}(t)$ being previously obtained from the
dynamics Eq. (\ref{eq:Master_LYP}) excluding the noise error, we
have shown the steady-state fidelity (at $\Omega_{r}t/2\pi=2500$)
as functions of the noise factors $\eta_{k}$ in Fig. \ref{Fig_Fidelity_vs_noise}.
It can be found that the steady-state fidelity is reduced by $\sim2\%$
for $\eta_{1}=0.05$ and $\eta_{3}=0.05$, corresponding to 5\% of
random fluctuations in the microwave driving intensity and the laser
driving strength, respectively. However, the steady-state fidelity
is less insensitive to random noise in the microwave detuning $(\sim\eta_{2})$
and the Rydberg-Rydberg interaction strength $(\sim\eta_{4})$. 

\section{Conclusion}

In conclusion, we have proposed the improved dissipation-assisted
scheme for preparing the two-atom singlet state under the Rydberg
antiblockade and the Lyapunov control. By appropriately selecting
the detuning and coupling strength between the microwave driving and
the level separation of the two clock states, the system converges
to the steady state faster than that with resonant microwave driving.
By implementing the Lyapunov control, the system can speed up the
convergence if the initial state involves the coherence between the
singlet state and the bright state. The improved scheme involving
the Lyapunov control is very efficient for a slow dissipative dynamics
and becomes less beneficial for a fast decaying system. The scheme
may be realized by the state of the art Rydberg experiments and can
be potentially generalized to multi-atom scenario.

\section*{ACKNOWLEDGMENTS}

L.-T.S., Z.-B.Y., H.W., and S.-B.Z. are supported by the National
Natural Science Foundation of China under Grants No. 11774058, No.
11674060, No. 11874114, and No. 11705030, the Natural Science Foundation
of Fujian Province under Grant No. 2017J01401, and the Qishan fellowship
of Fuzhou University. H.W. acknowledges particularly the financial
support by the China Scholarship Council for the academic visit to
the University of Nottingham.

\end{document}